\documentstyle[aps,epsfig,preprint]{revtex}
\tightenlines
\begin{document}
{\title{Microwave spectroscopy on a double quantum dot \\ 
with an on-chip Josephson oscillator}
\draft
\author{A.W. Holleitner, H. Qin, F. Simmel, B. Irmer, R.H. Blick$^*$, and
J.P. Kotthaus \\A.V. Ustinov$^{\dagger}$ and K. Eberl$^{\ddagger}$} 
\address{         Center for NanoScience and Sektion Physik,
Ludwig-Maximilians-Universit\"at, Geschwister-Scholl-Platz 1,
80539 M\"unchen, Germany. \\$^{\dagger}$  
Physikalisches Institut III, Universit\"at
Erlangen-N\"urnberg, Erwin-Rommel-Str. 1, \\
91058 Erlangen, Germany. 
\\$^{\ddagger}$ Max-Planck-Institut f\"ur Festk\"orperforschung,  
Heisenbergstr. 1, 70569 Stuttgart, Germany.         } 

\date{\today} 
\maketitle
\begin{abstract}
We present measurements on microwave spectroscopy on a double quantum dot
with an on-chip microwave source. The quantum dots are realized in the
two-dimensional electron gas of an AlGaAs/GaAs heterostructure and are
weakly coupled in series by a tunneling barrier forming an 'ionic'
molecular state. 
As a microwave source we employ a Josephson oscillator formed by a long
Nb/Al-AlO$_x$/Nb junction. 
We find photon assisted tunneling sidebands induced by the Josephson
oscillator and compare the results 
with those using an externally operated microwave source. 
\end{abstract} 
\newpage 

Microwave spectroscopy on quantum dots promises to probe the dynamics of these
few electron systems. Most of the experimental work conducted to date has
been focused on 
rather simple spectroscopic tools: a microwave signal is coupled
via an antenna or a stripline to the mesoscopic system under test. 
The photon-induced current through the dots is measured and allows
to probe the discrete states of the quantum system
directly~\cite{kouwenhoven94,blick1,oosterkamp1}. 
These results can be described by the Tien-Gordon theory~\cite{tien63}
originally developed for 
a superconductor tunnel junction and more recent theoretical
models~\cite{stafford96:1916,ivanov,grifoni,inarrea}. 
In order to reveal the dynamics of electrons confined in tunnel coupled dot
systems, more intricate spectroscopic 
tools are required. In the work by Nakamura{~\it et al.}~\cite{nakamura}
it was shown how to monitor a 
single tunneling Cooper pair in a superconducting tunnel junction
transistor in the time domain. This spurs 
interest in tracing electrons in coupled quantum dots, since in this case a
similar tunnel splitting of 
the discrete states was found~\cite{blick2,oosterkamp2}. 

Here we present an on-chip microwave oscillator integrated in a single
low-temperature setup 
with a coupled quantum dot structure. Integrating on-chip microwave sources
has the advantage of combining 
advanced spectrometers easily with mesoscopic devices. Furthermore, the
influence of black-body radiation is 
minimized, since all the electrical connections to the outside world are
essentially dc-lines and can thus be 
heavily filtered~\cite{adourian98,visscher99}. As a microwave source we
employ a long Josephson tunnel junction~(JTJ) 
with well-defined emission characteristics. These junctions radiate
commonly in the range from few GHz up to
600~GHz~\cite{Ust-PhysD:98}. Their typical radiation linewidth can be as small 
as $10^{-6}$ relative to
the emission frequency$~\cite{Kosh-self-pump:PRB97}$. Recently, this triggered rapid
progress in using these devices in
integrated sub-millimeter wave superconducting
receivers~\cite{Koshel:APL96}. The quantum dots are defined in 
a two-dimensional electron gas (2DEG) by electron beam written lateral
Schottky gates. The versatility of these 
devices is the possibility to tune the tunnel contacts in a wide range.
This allows the straight forward realization of tunnel 
coupled quantum dots ('covalent artificial molecule') or decoupled dots
('ionic artificial molecule')~\cite{oosterkamp2}.

In contrast to the excitation spectrum of real atoms or molecules, the
spectrum of single or even coupled quantum 
dots reveals a striking difference in the discrete level structure. For
quantum dots it has been shown in a whole 
variety of experiments that Kohn's theorem~\cite{kohn61}
prevails~\cite{pfann,dots1}. The theorem states that only 
the center-of-mass (CM) degree of freedom couples to a spatially homogenous
electromagnetic field. In previous 
studies of excitations in quantum dots by coupling radiation via antennas
only CM excitations were found ~\cite{oosterkamp2,dots1,fujisawa1}. 
Here we 'move' the radiation source 
close to the quantum dots, in an attempt to verify whether the inhomogeneity 
of the near-field radiation affects the electronic excitations. 
This is in close analogy to the probing of single molecules by scaning
near-field microscopy \cite{novotny96}.

The setup we used is shown in Fig.~1: the Si-chip with the
Nb/Al-AlO$_{x}$/Nb Josephson junction is glued on top 
of the quantum dot AlGaAs/GaAs-chip with photo resist. In the inset of Fig.~1
the quantum dot gate structure used in this work is depicted: Application of an
appropiate negative gate voltage defines two quantum dots in 
the 2DEG of the AlGaAs/GaAs heterostructure 
with an electron density of $1.7 \times 10^{15}$~m$^{-2}$ at 35~mK
(marked by the upper white circle in Fig.~1).
By variation of the voltages $V_{gA}$
and $V_{gB}$ applied to the plunger gates denoted 
in the inset of Fig.~1 the electron configuration of the double dot can 
be changed. Plotting the current through the system as a function of $V_{gA}$
and $V_{gB}$ results in the charging diagram of Fig.~2. 
Subsequently, we obtain charging energies of the individual dots of
$E_{C}^{A} = {e^2}/{2C_{\Sigma}} = 220 ~\mu$eV 
and $E_C^B = 205 ~\mu$eV, respectively.
Thus the 'electronic' radii are $r_A = 400 $~nm and $r_B = 430 $~nm. 
For the absolute number of electrons in each dot we find: $n_A=~850 $ 
and $n_B =~980 $. Hence, the dots are rather classical systems in which the
resolution of discrete single particle energies due to the confinement
potential is not expected. In the current measurements the interdot coupling is chosen 
to be weak ($C_{A-B} = 2$~aF), i.e. we see a hexagonal 
array of points for the charging diagram~\cite{livermore} 
which corresponds to the 'ionic' coupling limit. 
All results were obtained in the linear regime (drain/source bias $V_{ds}=19 ~\mu$V).

In an earlier setup we focused on defining 
a weak link with a modified atomic force microscope tip directly into the
Al-contacts forming the Schottky 
gates~\cite{irmer1,irmer2}. This is of great advantage for probing the
microwave response of quantum dots in the absolute
near-field limit, since the photon source is located only 100~nm apart from the
tunnel barriers. However, these weak links 
radiate with very large radiation linewidth. Hence, we choose here to employ a
well characterized JTJ oscillator placed on top
of the quantum dots. The JTJ we use in these measurements is an overlap
Nb/Al-AlO$_x$/Nb long Josephson junction 
with dimensions $20 \times 400 ~\mu$m$^2$ (width$\times$ length). In the
measurements presented, tuning of the frequency 
was possible by varying the JTJ critical current with the applied magnetic
field and selecting the
appropriate bias point at a resonant state. When operated in the flux-flow
regime, the JTJ radiation 
frequencies $f = 2 e V / h$ are on the order of 100 --~500~GHz. This is
above the Coulomb energy for this particular 
quantum dot. We employ a finite magnetic field to operate the junction at a
Fiske step of the current/voltage 
characteristic, a self-resonant state. The fundamental cavity resonance
frequency of the junction is $f=\bar{c}/(2\ell)$, 
where $\bar{c}$ is the Swihart velocity and $\ell$ is the junction length.

In order to characterize the 
microwave response of the coupled quantum dots we 
first studied them with a conventional microwave source. The radiation was
then transduced by 
an antenna upon the gate structure of the quantum dots.
The charging diagram of the double dot with microwave excitation is shown
as a grayscale plot 
in Fig.~3. The radiation of the far-field source is fixed at 10~GHz. The sidebands are
found on only one side, depending on the 
tuning of the barrier transmission coefficients. In measurements on a similar 
two-dot device we obtained symmetrically as well as
asymmetrically induced sidebands in 
the charging diagram~\cite{qin99}. However, since we are
interested in the alteration of the microwave 
coupling by the JTJ we chose deliberately to maintain the tuning with only one
sideband clearly appearing in the inset of Fig.~3. 
This finally ensures that we can directly compare the measurements
with the conventional microwave source and the on-chip source. The inset shows one 
of the typical traces from the grayscale plot with the sidebands induced by the
frequency dependent absorption. 
The induced sidebands are marked by arrows -- the peak height modulation is
due to the specific trace taken out of 
the charging diagram (marked by the dashed line). A cut along one of the two
periodic resonance lines
would yield peaks of similar amplitude. The net pumping of electrons leads
to a reduction of the absolute current 
value down to only some 100~fA. Also the noise floor is slightly enhanced
by coupling the radiation.

When the on-chip Josephson oscillator is operated as a source with a
typical emission frequency 
of $f \cong 10$~GHz, we observe charging diagrams as the one shown in
Fig.~4. The JTJ emission frequency 
was determined by taking the $IV$-characteristics. Biasing the JTJ with a
current of $I = 1 ~\mu$A we 
are able to detect sidebands which resemble the ones induced by the
far-field source (compare insets of Fig.~3 and~4).
The power emitted is then on the order of $P_{\rm dc} \sim
20~\mu$V$\times\,1\,\mu$A$~\approx20\,$pW, where $20~\mu$V is the 
voltage drop over the JTJ  at 10 GHz. Moreover, the peak broadening is almost
identical to the one determined before. 
As seen, the resonances of the current (marked by lines in the plot) and
the induced sidebands 
(marked by arrows) possess a long term stability. Since the observed
resonances for the on-chip source
as well as the far-field source are almost identical, we conclude that the
photon absorption process 
only depends on the shape of the local electrostatic potential.   
Varying the frequency of the radiation for such a tunnel coupled dot system
results in the well-known linear 
relation between the position of the sideband relative to the ground state and
the frequency for the 
case of weak coupling~\cite{oosterkamp2,qin99}.

In summary, we find photon-assisted tunneling in a weakly coupled double
quantum dot, induced by an
on-chip source. This source is realized as a long Josephson junction
placed on top of the chip carrying 
the double dot. We find nearly identical coupling of radiation with the
on-chip source and with that 
of the far-field source. We conclude from this comparison that the photon
absorption process depends only 
on the local electrostatic environment of the quantum dots. Furthermore,
this result confirms that 
Kohn's theorem is valid in the near-field regime, as long as it provides a
spatially homogenous radiation
field across the excited electronic system. This we expect to change when
a Josephson junction is embedded in the dot's gate structure. 

We like to thank T. Klapwijk, S.J. Allen, and D. Grundler for extended
discussions.  
This work was funded in part by the Deutsche Forschungsgemeinschaft
(DFG), the Defense Advanced Research Projects Agency (DARPA) 
and H. Qin likes to thank the Volkswagen-Stiftung for support. 

$^*$ Corresponding author: Robert H. Blick, Center for NanoScience and
Sektion Physik, 
Ludwig-Maximilians-Universit\"at, Geschwister-Scholl-Platz 1, 80539
M\"unchen, Germany. \\
Electronic mail: {\it robert.blick@physik.uni-muenchen.de}



\figure{Fig.~1: Top view of the circuit: The quantum dots are located in the
center of 
the Hall-bar (marked by the upper white circle). An AFM micrograph of the
gates forming the double 
dots is seen in the inset on the upper right side (gA, gB denote the plunger gates 
operated in the measurements). The Josephson oscillator is placed on 
top of the quantum dot chip and glued to it with photo resist. The lower
white circle indicates the 
position of the junction itself. Radiation is coupled through the GaAs
substrate to the quantum 
dots (see text for details).  
   }
\label{one}

\figure{Fig.~2: Charging diagram of the double quantum dot as a grayscale
plot without
microwave radiation (white: $I < 0$~pA, black: $I > 0.5$~pA).  
A small forward bias of $V_{ds} = 19 ~\mu$V is applied to monitor the
current. The two dots are weakly coupled by the tunnel barrier and 
produce a periodic lattice with $E_C^A$ and $E_C^B$ denoting the charging
energies, as marked
by the diamond. The two gate voltages $V_{gA}$ and $V_{gB}$  span 
the charging diagram.  Inset gives a line plot of the direct-current through the
double dot. The range of voltage $V_{gB}$ in this graph coincides with the broken 
line in the gray-scale plot.      
} 
\label{two}

\figure{Fig.~3: Charging diagram with an identical scan range as in
Fig.~2, but under 
microwave radiation by a far-field source at 10~GHz. The sidebands are
found on only one side, 
depending on the tuning of the tunnel barriers. Inset: Photo current with
the sidebands induced by 
the frequency dependent absorption. Pumping of the electrons leads to a
reduction of the absolute current value.  The gray lines denote the range of  voltage
$V_{gB}$ which correlates to the frequency of $f = 10$~GHz on this axis.
}
\label{three}

\figure{Fig.~4: Charging diagram under microwave radiation with the on-chip
Josephson oscillator as 
a source ($f \cong 10$~GHz). Inset gives the photon-induced current through
the dot 
system with the photon assisted tunneling resonances. The peak heights and
positions are 
identical to the plots obtained with the far-field source (compare Fig.~3). This
similar pattern indicates 
that the coupling of radiation of near- and far-field only depends on the
local electrostatic environment.
}
\label{four}


\begin{thebibliography}{10}

\bibitem{kouwenhoven94} L.P. Kouwenhoven, S. Jauhar, J. Orenstein, P.L.
McEuen, 
Y. Nagamune, J. Motohisa, and H. Sakaki, Phys. Rev. Lett. {\bf 73}, 3443 (1994).

\bibitem{blick1} R.H. Blick, R.J. Haug, D.W. van der Weide, K. von
Klitzing, and K. Eberl, Appl. Phys. Lett. {\bf 67}, 3924 (1995).

\bibitem{oosterkamp1} T.H. Oosterkamp, L.P. Kouwenhoven, A.E.A. Koolen, N.C.
van der Vaart, and C.J.P.M. Harmans, Phys. Rev.Lett. {\bf 78}, 1536 (1997).

\bibitem{tien63} P.K. Tien and J.P. Gordon, Phys. Rev. {\bf 129}, 647 (1963).

\bibitem{stafford96:1916} C.A. Stafford and N.S. Wingreen, Phys. Rev.
Lett. {\bf 76}, 1916 (1996).

\bibitem{ivanov} T. Ivanov, Phys. Rev. B {\bf 56}, 12339 (1997); T.H. Stoof 
and Yu.V. Nazarov, Phys. Rev. B {\bf 53}, 1050 (1996).

\bibitem{grifoni} M. Grifoni and P. H\"anggi, Physics Reports {\bf 304}, 
229 (1998).

\bibitem{inarrea} J. Inarrea, G. Platero, and C. Tejedor, Phys. Rev. B
{\bf50}, 4581 (1994).

\bibitem{nakamura} Y. Nakamura, Yu.A. Pashkin, and J.S. Tsai, Nature
{\bf398}, 786 (1999).

\bibitem{blick2} R.H. Blick, D.W. van der Weide, R.J. Haug, K. Eberl, Phys.
Rev. Lett. {\bf 81}, 689 (1998). 

\bibitem{oosterkamp2} T.H. Oosterkamp, T. Fujisawa, W.G. van der Wiel,
K. Ishibashi, R.V. Hijman, S. Tarucha, 
and L.P. Kouwenhoven, Nature {\bf 395}, 873 (1998).


\bibitem{adourian98} A.S. Adourian, S. Yang, and R.M. Westervelt, K.L.
Chapman and A.C. Gossard, J. Appl. Phys. {\bf 84}, 
5808 (1998).

\bibitem{visscher99} E.H. Visscher, D.M. Schraven, P. Hadley, and J.E.
Mooij, cond-mat/9904382.

\bibitem{Ust-PhysD:98} A.V. Ustinov. Physica D {\bf 123}, 315-329 (1998).

\bibitem{Kosh-self-pump:PRB97} V.P. Koshelets, S.V. Shitov, A.V. Shukin,
L.V. Filippenko, J. Mygind, 
and A.V. Ustinov, Phys. Rev. B  {\bf 56},5572 (1997).

\bibitem{Koshel:APL96} V.P.~Koshelets, S.V.~Shitov, L.V.~Filippenko,
A.M.~Baryshev, 
H.~Golstein, T.~de~Graauw, W.~Luinge, H.~Schaeffer, and H.~van~de~Stadt,  
Appl. Phys. Lett. {\bf 68}, 1273 (1996).

\bibitem{kohn61} W. Kohn, Phys. Rev. {\bf 123}, 1242 (1961). 

\bibitem{pfann} D. Pfannkuche and S.E. Ulloa, Phys. Rev. Lett. {\bf
74}, 1194 (1995); 
J. Weis, R.J. Haug, K. von Klitzing, and K. Ploog, Phys. Rev. Lett. {\bf
71}, 4019 (1993).

\bibitem{dots1} C.H. Sikorski and U. Merkt, Phys. Rev. Lett. {\bf 62},
2164 (1989); 
T. Demel, D. Heitmann, P. Grambow, and K. Ploog, Phys. Rev. Lett.  {\bf
64}, 788 (1990); 
U. Merkt, Phys. Rev. Lett. {\bf 76}, 1134 (1996);
P. Bakshi, D.A. Broido, and K. Kempa, Phys. Rev. B {\bf 42}, 7416 (1990). 


\bibitem{fujisawa1} T. Fujisawa and S. Tarucha, Superlattices and 
Microstructures {\bf21}, 247-254 (1997).

\bibitem{novotny96} L. Novotny, Appl. Phys. Lett. {\bf 69}, 3806 (1996).

\bibitem{livermore} C. Livermore, C.H. Crouch, R.M. Westervelt, K.L.
Campman, and A.C. Gossard, 
Science {\bf 274}, 1332 (1996).

\bibitem{irmer1} B. Irmer, R.H. Blick, F. Simmel, W. G\"odel, H. Lorenz, 
and J.P. Kotthaus, Appl. Phys. Lett. {\bf 73}, 2061 (1998). 

\bibitem{irmer2} B. Irmer, F. Simmel, R.H. Blick, H. Lorenz, J.P. Kotthaus, 
M. Bichler, and W. Wegscheider, Superlattices and Microstructures {\bf 25}, 785 (1999).

\bibitem{qin99} H. Qin, R.H. Blick, F. Simmel, A.W. Holleitner, J.P.
Kotthaus, and K. Eberl, to be submitted (1999).

\end{thebibliography}
\end{document}